\documentclass[%
reprint,
superscriptaddress,
 amsmath,amssymb,
 aps,
pra,
]{revtex4-2}
\usepackage{upgreek}
\usepackage{graphicx}
\usepackage{dcolumn}
\usepackage{bm}
\usepackage{hyperref}
\usepackage[mathlines]{lineno}

\begin{document}


\title{Integration of a GaAs-based nanomechanical phase shifter with quantum-dot single-photon sources}

\author{Celeste Qvotrup}
\thanks{These two authors contributed equally.}
\affiliation{Center for Hybrid Quantum Networks (Hy-Q), Niels Bohr Institute, University of Copenhagen, Blegdamsvej 17, Copenhagen, 2100, Denmark.}

\author{Ying Wang$^\dag$}
\thanks{These two authors contributed equally.}
\affiliation{Center for Hybrid Quantum Networks (Hy-Q), Niels Bohr Institute, University of Copenhagen, Blegdamsvej 17, Copenhagen, 2100, Denmark.}

\author{Marcus Albrechtsen}
\affiliation{Center for Hybrid Quantum Networks (Hy-Q), Niels Bohr Institute, University of Copenhagen, Blegdamsvej 17, Copenhagen, 2100, Denmark.}
\author{Rodrigo A. Thomas}
\author{Zhe Liu}
\affiliation{Center for Hybrid Quantum Networks (Hy-Q), Niels Bohr Institute, University of Copenhagen, Blegdamsvej 17, Copenhagen, 2100, Denmark.}
\author{Sven Scholz}
\affiliation{Lehrstuhl f{\"u}r Angewandte Festk{\"o}rperphysik, Ruhr-Universit{\"a}t Bochum, Universit{\"a}tsstrasse 150, D-44780 Bochum, Germany}

\author{Arne Ludwig}
\affiliation{Lehrstuhl f{\"u}r Angewandte Festk{\"o}rperphysik, Ruhr-Universit{\"a}t Bochum, Universit{\"a}tsstrasse 150, D-44780 Bochum, Germany}

\author{Leonardo Midolo}
\thanks{Email to: ying.wang@nbi.ku.dk; midolo@nbi.ku.dk}
\affiliation{Center for Hybrid Quantum Networks (Hy-Q), Niels Bohr Institute, University of Copenhagen, Blegdamsvej 17, Copenhagen, 2100, Denmark.}
\date{\today}

\begin{abstract}
We demonstrate a small-footprint electromechanical phase shifter with a compact active length of $10\:\upmu \text{m}$ fabricated on a suspended GaAs membrane, offering versatile integration with quantum-dot single-photon sources. The phase shifter is based on a slot-mode waveguide, whose slot width can be controlled by electrostatic forces, enabling a large effective refractive index change  $\Delta n_\text{eff} > 0.1$. We observe up to 3$\pi$ phase modulation with 10.6 \text{V} applied bias, and a figure of merit $V_{\pi}L = 5.7\cdot 10^{-3} \text{V}\cdot \text{cm}$. Integration with a Mach-Zehnder interferometer (MZI) further allows routing of single photons with up to 24 \text{dB} extinction ratio at cryogenic temperatures. This device enables advanced manipulation of quantum emitters and the realization of reconfigurable quantum photonic integrated circuits. 
\end{abstract}

\maketitle
\section{Introduction}
Recent advancements in the growth and processing of gallium arsenide (GaAs) photonic integrated circuits (PICs) have opened up promising avenues to the realization of quantum photonic integrated processors where single-photon emitters, routers, and detectors can be seamlessly integrated within a single chip \cite{dietrich2016gaas,Uppu2021}. By combining self-assembled indium arsenide (InAs) quantum dots (QDs), and programmable PICs, photon losses can be greatly suppressed, thereby boosting the performance of many quantum applications, such as quantum key distribution  and quantum simulation \cite{maring2024versatile}, which frequently require hybrid or modular architectures \cite{morrison2023single,zahidy2024quantum,maring2024versatile,wang2023deterministic}. 
Achieving the monolithic integration of all the functionalities in a single material platform poses considerable challenges. While both quantum emitters \cite{uppu2020scalable, tomm2021bright,papon2023independent} and superconducting single-photon detectors \cite{sprengers2011waveguide,reithmaier2013optimisation} have been realized in GaAs, the implementation of scalable reconfigurable devices in this platform is still undeveloped. A compact, tunable phase shifter that operates at cryogenic temperatures is crucial for quantum photonic integrated circuits (PICs), as it allows for the programmable transformation of photonic qubits and the construction of logical gates \cite{bogaerts2020programmable}.

Thermo-optic phase shifters, which modulate the phase by changing the refractive index through temperature variations, are preferred in silicon \cite{harris2014efficient} and silicon nitride \cite{nejadriahi2021efficient}  PICs due to their low loss and ease of fabrication. However, this method is less suitable for the GaAs platform because of factors such as the poor thermal conductivity, the near-zero thermo-optic coefficient at cryogenic temperatures, and the large device footprints. GaAs electro-optic phase shifters have been realized at telecom wavelengths with GHz-scale modulation rates and CMOS-compatible driving voltages \cite{wood1984high}, owing to the high refractive index of GaAs and moderate Pockels coefficients \cite{thomaschewski2022pockels}. Nevertheless, this approach faces challenges at the 930 nm wavelength, where QDs emit, due to Franz-Keldysh electroabsorption leading to excessive insertion loss \cite{midolo2017electro, wang2021electroabsorption}. 
An ideal alternative solution to programmable circuits is offered by nano-opto-electro-mechanical systems (NOEMS). In NOEMS, light control is achieved by mechanically adjusting the geometry of the guided mode, thereby altering its optical phase \cite{midolo2018nano}. In a previous work, we reported a gap-variable four-ports directional coupler controlled by NOEMS on GaAs which demonstrated the compatibility of nanomechanical motion with QDs at cryogenic temperatures \cite{papon2019nanomechanical}. 
A variety of reconfigurable NOEMS phase shifters, designed using hybrid photonic-plasmonic modes, coupled-waveguide modes, and slot waveguide modes, with sub-dB insertion loss, compact footprints, and low driving power have been demonstrated on silicon and silicon nitride platforms at room temperature \cite{van2012ultracompact,haffner2019nano,edinger2021silicon,baghdadi2021dual}. In silicon photonics, slot waveguides \cite{almeida2004guiding} electro-mechanical devices have been reported, featuring large phase shifts \cite{grottke2021optoelectromechanical}  and excellent performance at cryogenic temperatures \cite{beutel2022cryo}. 

In this work, we demonstrate a reconfigurable slot-mode waveguide NOEMS phase shifter interfaced with a QD single-photon source on suspended GaAs waveguides. The resulting device enables a low half-wave-voltage length product, $V_{\pi}L = 5.7\cdot 10^{-3}$ $ \text{V}\cdot \text{cm}$, with a compact active length of 10 $\upmu$m and an average insertion loss of $-2.5$ dB. The reported phase shifter in GaAs is the missing element to realize full nanomechanical unitary transformations of dual-rail encoded photonic qubits integrated with single-photon sources.  

\begin{figure}[ht]
\includegraphics{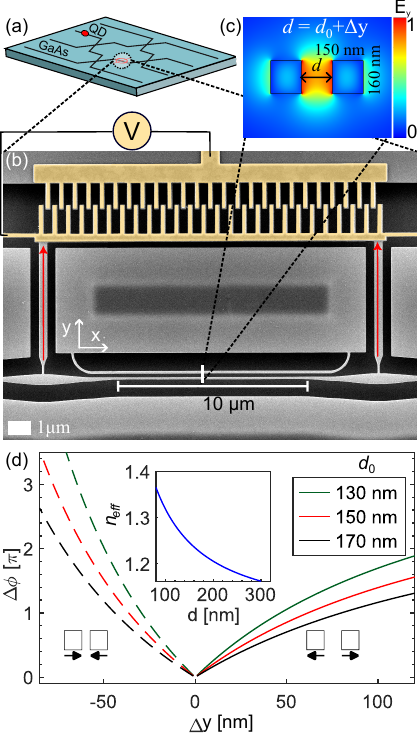}
\caption{\label{fig:1} Design and working principle of the NOEMS phase shifter. (a) Sketch of a GaAs photonic integrated circuits with QDs and reconfigurable phase shifters. (b) Scanning electron microscope image (SEM) of the fabricated nanomechanical phase shifter comprising a 10-$\upmu$m-long slot-mode waveguide driven by an electrostatic comb drive (highlighted in yellow).  (c) Finite element simulation of the pseudo-transverse electric (TE) mode in the slot waveguide. (d) Theoretical phase shift $\Delta\phi$ as a function of the displacement $\Delta y$ for the slot-mode waveguide for three different initial slot widths $d_0$. Inset: Effective refractive index as a function of the slot width calculated at the wavelength of 950 nm. }
\end{figure}

\section{Working principle and design of the phase shifter}

The phase shifter is realized on a suspended 160-nm-thick GaAs membrane with embedded self-assembled QDs and nanophotonic waveguides, as illustrated in Figure \ref{fig:1}(a). Fig. \ref{fig:1}(b) shows a scanning electron microscope (SEM) image of the fabricated NOEMS phase shifter. The device features a slot waveguide formed by two 150-nm-wide single-mode waveguides, where one of them is connected to a Cr/Au (10/160 nm) electrostatic comb drive (highlighted in yellow) deposited on top of GaAs \cite{papon2019nanomechanical}. The length $L$ of the slot-mode waveguide, responsible for phase modulation, is 10 $\upmu$\text{m}, the longest that could be reliably fabricated without adding lossy support tethers and without incurring mechanical failures. The waveguide is designed to support a single transverse electric (TE) mode, shown in Fig. \ref{fig:1}(b) with low effective refractive index $n_\text{eff}\simeq1.2$. The phase modulation $\Delta \phi =\Delta n_\text{eff}kL$ is achieved via controlling \text{$n_\text{eff}$} by driving the electrostatic actuator with a voltage source. Given the short length of the device, at least 5\% modulation of the effective refractive index is required to achieve a $\pi$-shift. Figure \ref{fig:1}(d) shows the calculated phase shift as a function of the displacement $\Delta y$ for different initial slot widths $d_0$, obtained from finite element simulations of the waveguide mode. Taking advantage of the large nonlinear dependence of $n_\text{eff}$ on the slot width, shown in the inset of Fig. \ref{fig:1}(d), up to 3$\pi$ phase shift is theoretically achievable when reducing the slot width (negative displacements) by roughly 60 nm. The plot also illustrates that operating the device in \emph{push}-mode is significantly more efficient in modulating the phase than \emph{pull}-mode (i.e., for positive displacements). Moreover, wider initial gaps $d_0$ (up to 200 nm) not only enable sufficient phase shift but also improve fabrication yield, ensuring reliable integration into photonic circuits and scalability.
The mechanical displacement of the combs under an external applied bias voltage $V$ follows  
\begin{equation}
    d(V) = d_0-\eta V^2, 
    \label{eq_gapRT}
\end{equation}
with $\eta$ representing the electro-mechanical transduction efficiency given by the geometry of the comb drive and the stiffness of the mechanical structure \cite{legtenberg1996comb}. We designed the suspension tethers holding comb drive and waveguides using finite element modeling to balance in-plane and out-of-plane device stiffness and maximize $\eta$. 

\begin{figure}[hbt]
\centering
\includegraphics{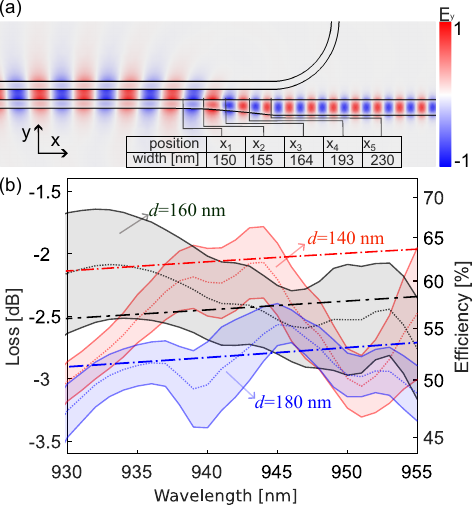}
\caption{\label{fig:2}
(a) Finite-element simulation of the electric field in the 2-$\upmu$m-long taper used for slot waveguide coupler. The table indicates the waveguide width as a function of position where $x_n-x_{n-1} = 400$ nm. (b) Wavelength-dependent insertion loss of the slot waveguide phase shifter for three different slot widths $d$. Dash-dotted lines show theoretical losses, while banded lines show measured losses, with the upper and lower limit of the band corresponding to measurement uncertainties.}
\end{figure}

To keep losses at a minimum, the waveguides are first tapered out to 600 nm width, to avoid scattering losses at the position of the tether, then adiabatically reduced to 150 nm. 
This is achieved with a piece-wise linear taper, which is designed to gradually reduce the effective index along the propagation direction. 
Fig.~\ref{fig:2}(a) shows the electric field distribution across the taper transition calculated via three-dimensional finite element analysis. The calculated losses are shown as a function of wavelength in Fig.~\ref{fig:2}(b) (dash-dotted lines). We observe that each taper contributes to 1--1.5 dB in the typical 930-950 nm wavelength range of QD emission leading to a total insertion loss of between $-2$ and $-3$ dB per device. Moreover, losses tend to reduce at smaller slot widths which could be explained by the higher confinement and effective index, facilitating an adiabatic transition. 
To compare with simulations, we fabricated a large amount of nominally identical phase shifters and compared the transmission across 0, 1, 2, or 3 devices connected in series at different wavelengths in the transmission range of our grating couplers \cite{zhou2018high}. The resulting loss per device matches the prediction of simulations with the measurement uncertainties. Further improvements of the adiabatic couplers are possible, for example considering different optimization strategies and more advanced taper geometries. 

\section{Characterization of the phase shifter}
\begin{figure}[ht]
\includegraphics{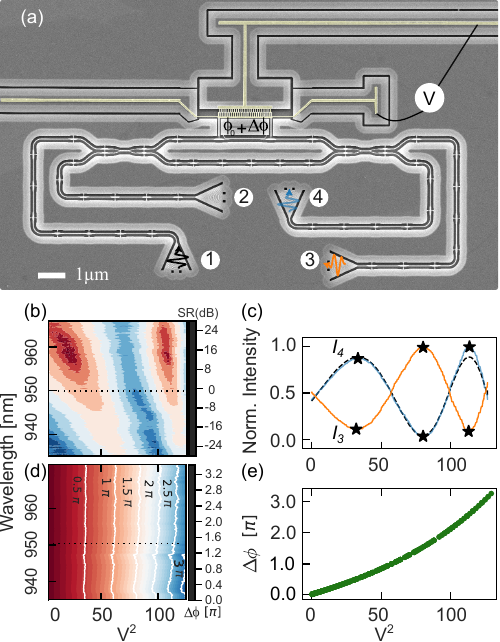}
\caption{\label{fig:3}
Characterization of the NOEMS phase shifter at room temperature.
(a) A schematic illustrating the experimental operation. A continuously tunable laser is coupled into the MZI via grating 1 or 2, with transmitted (reflected) light coupled out through grating 3 (4). An external voltage bias is applied to the phase shifter to modulate the phase. 
(b) Coupling-independent split ratio (SR) dynamics are observed by sweeping the wavelength and voltage bias.
(c) Intensity modulation at the output port 3 (orange) and 4 (blue) at a wavelength of 950 \text{nm}. The dash line is a fit to the intensity modulation assuming a constant visibility.
(d) Phase modulation \text{$\Delta \phi$} as a function of wavelength and voltage bias and (e) cross-section at 950 nm.
}
\end{figure}

To test the NOEMS phase shifter, a Mach-Zehnder Interferometer (MZI) was designed, as shown in Fig. \ref{fig:3}(a). 
Two directional couplers are used as input and output beam splitters, designed to a 3 dB splitting ratio (SR) at the wavelength of 950 nm. Light is coupled into the chip from an objective into focusing grating couplers \cite{zhou2018high} whose cross-polarized orientation on the chip enable suppressing scattered light in transmission measurements.  

The device is characterized in vacuum and at room temperature first to evaluate the phase shift performance. A tunable continuous-wave laser is used to probe the devices and the collected light is recorded using Si avalanche photodiodes. Transmission measurements between the input ports (labeled as 1 and 2 in Fig.~\ref{fig:3}(a)) and the output ports (3, 4) are taken to determine the intensity response \text{$I_{ij}$} between port $i$ and port $j$. This approach allows for precise evaluation of the intrinsic performance of the phase shifter, independent of grating coupling efficiency and collection efficiency between the output ports. In Fig.~\ref{fig:3}(b) the total coupling-independent SR of the MZI, defined as $\sqrt{\frac{I_{13}I_{24}}{I_{14}I_{23}}}$ is plotted as a function of wavelength and of the applied voltage squared (proportional to the nanomechanical displacement of the slot waveguide). 
A clear oscillatory response to the applied bias is observed at all wavelengths. 
A global maximum (minimum) SR of $23.4\pm 0.5$ ($-23.7\pm0.5$) \text{dB} is measured at 937 (961) nm, respectively. SR values decrease as the wavelength shortens, suggesting the presence of asymmetric and wavelength-dependent loss mechanisms in the MZI. These are likely caused by a combination of imbalanced splitting ratios in the directional couplers, as well as a loss difference between the two arms of the MZI.
The normalized intensity plot, taken at a single wavelength of 950 nm, is shown in Fig. \ref{fig:3}(c)). 
We model the transmission across the MZI as follows:
\begin{align}
  I_{\text{out},4(3)} = \frac{I_0}{2}\left(1\pm\nu_\text{4(3)} \cos(\phi_0 +\Delta \phi)\right)
  \label{eq4}
\end{align}
where $\nu_\text{4}$ and $\nu_\text{3}$ denote the (wavelength-dependent) interference visibility at port 4 and 3, respectively. 
From the extrema of the fringes (marked with stars in Fig.~\ref{fig:3}(c)), we obtain \text{$\nu_\text{4,1(2)} = 0.80(0.84)$}  and\text{ $\nu_\text{3,1(2)} = 0.92(0.93)$}, where 1(2) denotes the first (second) extrema set, respectively. The small increase of visibility at higher bias is consistent with the theoretical prediction that the phase shifter insertion loss decreases with the slot width (cf. Fig.~\ref{fig:2}(b)).
The oscillatory response of the intensity can be used to directly compute the relative phase shift $\Delta \phi (V)$ at each voltage value. The result is shown in Fig. \ref{fig:3}(d) for all wavelengths and (e) at 950 nm.  The white iso-lines indicate the amount of phase shift relative to zero bias achieved at each wavelength, confirming the broadband operation of the device and almost near-identical performance over a 30 nm bandwidth.
A \text{$\Delta \phi$} of 3\text{$\pi$} is achieved with this compact phase shifter, requiring an applied potential of 10.6 V. 
We extract a figure of merit \text{$V_{\pi}$L} of \text{$5.7 \cdot 10^{-3}\text{V}\cdot\text{cm}$} considering the voltage ($V_\pi$=5.7 V) required to obtain a $\pi$ phase shift from the unbiased device. This value is further improved to $V_\pi = 1.7$ V when considering a biasing voltage of $V_0= 8.9$ V. Comparing this result to the simulated phase shift of Fig.~\ref{fig:1}(d), we estimate that the maximum mechanical displacement is $\sim80$ nm from an initial gap of 150 nm, corresponding to an electro-mechanical transduction at room temperature $\eta\simeq 0.6$ nm/V$^2$.

\section{Single-photon routing}
\begin{figure}[h]
\includegraphics{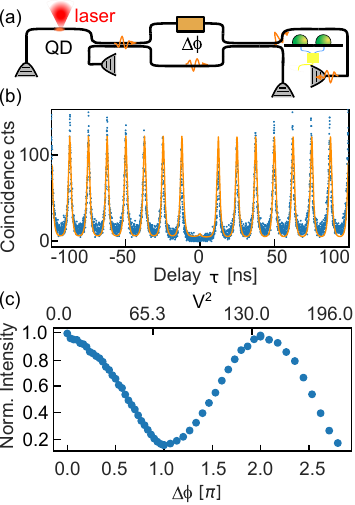}
\caption{\label{fig:4} 
(a) Schematic of the NOEMS photon router based on a tunable MZI. QDs are excited on the input arm and analyzed at one output. (b) Auto-correlation histogram for a QD emitting at 923 \text {nm} showing clear anti-bunching at zero-time delay. (c) Intensity modulation of the single-photon signal as a function of the applied squared voltage and corresponding phase shift.
}
\end{figure}
The performance of the NOEMS phase shifter at cryogenic temperature is characterized using InAs QDs operated at 10 \text{K} as a source of single photons. We employ the MZI as a photon router as illustrated in Figure \ref{fig:4} (a). The QDs are excited using a 800 nm above-band pulsed diode laser. The collected signal is filtered by a diffraction grating to select individual QD transitions. To confirm the single-photon nature of the emitter, an auto-correlation measurement $g^{(2)}(\tau)$ is performed on a single excitonic transition at 923 \text{nm}. The result is shown in Fig.~\ref{fig:4}(b), exhibiting a clear anti-bunching dip at zero delay ($g^{(2)}(0)=(0.007\pm0.003$), which indicates the efficient suppression multi-photon component from laser leakage and re-excitation processes. The observed increase in peak intensity over short time scale is due to artifacts in the photon counter instrumentation and the value of $g^{(2)}(0)$ is obtained by normalizing the data at long values of $\tau\simeq1$ ms. 

Fig.~\ref{fig:4}(c) shows the intensity modulation of the same QD emission line at 923 nm. A higher applied bias is required at cryogenic temperature compared to room temperature. At the highest applied bias of 14 V a $2.7\pi$ phase shift is achieved. We estimate that the electro-mechanical transduction $\eta$ reduces by approximately two thirds of the room temperature value at 10 K. 
We attribute this behavior to two main factors: (i) at cryogenic temperature GaAs and the Cr/Au metal contacts exhibit thermal compression at different rates, owing to the different coefficients of thermal expansion, which causes the whole suspended structure to displace out of plane. This could lead to a partial disengagement of the NOEMS electrostatic comb drive, resulting in a lower actuation force from the electrodes. (ii) Changes in the elastic modulus of Au, which becomes significantly higher at cryogenic temperatures, making the structure stiffer \cite{Noel2016GoldMembranes}.

\section{Conclusions}
We presented a cryogenically-compatible NOEMS phase shifter on a GaAs platform, a critical component for photonic quantum information processing. This platform combines high-purity quantum emitters with the capability to perform on-chip optical routing. The phase shifter leverages the reconfigurable coupling of slot waveguide modes, achieving an effective index change on the order of \text{$\Delta n_{\text{eff}}$} $\approx $0.1 with nanoscale displacement (tens of nanometers) of the slot waveguides. 
Importantly the device performance remains robust at cryogenic temperatures, critical for applications involving quantum emitters such as InAs quantum dots.
Although the phase shifter introduces a modest insertion loss of $<3$ \text{dB}, primarily attributed to scattering and surface charge absorption, these losses can be mitigated through surface passivation techniques \cite{Guha:17} and improved taper designs can be made to reduce losses even further. 
Integrating the phase shifter with gap-variable directional couplers \cite{papon2019nanomechanical}, would enable realizing low-power arbitrary 2x2 unitary transformations \cite{reck1994experimental} and potentially scale up further to large networks of multi-mode interferometers \cite{taballione202320}, with applications in boson sampling, quantum simulation \cite{sparrow2018simulating}, and machine learning \cite{bandyopadhyay2024single}. Additionally, the integration of quantum emitters and phase shifters in the same chip can be employed to fine-tune light-matter interaction, for example by tuning the local density of optical states and enabling remote control of spontaneous emission\cite{companion} or controlling the coupling between multiple emitters \cite{tiranov2023collective}.  
By advancing the scalability and compatibility of cryogenic photonic platforms, this device marks an important step forward in the realization of integrated quantum photonic architectures.

\begin{acknowledgments}
The authors ackowledge Jasper van der Weert for assistance in measuring the device loss. We acknowledge funding from the European Research Council (ERC) under the European
Union’s Horizon 2020 research and innovation program (No. 949043, NANOMEQ), the Danish National
Research Foundation (Center of Excellence “Hy-Q,” grant number DNRF139), Styrelsen for Forskning og
Innovation (FI) (5072- 00016B QUANTECH), BMBF QR. N project 16KIS2200, QUANTERA BMBF EQSOTIC project 16KIS2061, as well as DFG excellence cluster ML4Q project EXC 2004/1.
\end{acknowledgments}

\bibliography{bibliography}

\begin{thebibliography}{37}%
\makeatletter
\providecommand \@ifxundefined [1]{%
 \@ifx{#1\undefined}
}%
\providecommand \@ifnum [1]{%
 \ifnum #1\expandafter \@firstoftwo
 \else \expandafter \@secondoftwo
 \fi
}%
\providecommand \@ifx [1]{%
 \ifx #1\expandafter \@firstoftwo
 \else \expandafter \@secondoftwo
 \fi
}%
\providecommand \natexlab [1]{#1}%
\providecommand \enquote  [1]{``#1''}%
\providecommand \bibnamefont  [1]{#1}%
\providecommand \bibfnamefont [1]{#1}%
\providecommand \citenamefont [1]{#1}%
\providecommand \href@noop [0]{\@secondoftwo}%
\providecommand \href [0]{\begingroup \@sanitize@url \@href}%
\providecommand \@href[1]{\@@startlink{#1}\@@href}%
\providecommand \@@href[1]{\endgroup#1\@@endlink}%
\providecommand \@sanitize@url [0]{\catcode `\\12\catcode `\$12\catcode `\&12\catcode `\#12\catcode `\^12\catcode `\_12\catcode `\%12\relax}%
\providecommand \@@startlink[1]{}%
\providecommand \@@endlink[0]{}%
\providecommand \url  [0]{\begingroup\@sanitize@url \@url }%
\providecommand \@url [1]{\endgroup\@href {#1}{\urlprefix }}%
\providecommand \urlprefix  [0]{URL }%
\providecommand \Eprint [0]{\href }%
\providecommand \doibase [0]{https://doi.org/}%
\providecommand \selectlanguage [0]{\@gobble}%
\providecommand \bibinfo  [0]{\@secondoftwo}%
\providecommand \bibfield  [0]{\@secondoftwo}%
\providecommand \translation [1]{[#1]}%
\providecommand \BibitemOpen [0]{}%
\providecommand \bibitemStop [0]{}%
\providecommand \bibitemNoStop [0]{.\EOS\space}%
\providecommand \EOS [0]{\spacefactor3000\relax}%
\providecommand \BibitemShut  [1]{\csname bibitem#1\endcsname}%
\let\auto@bib@innerbib\@empty
\bibitem [{\citenamefont {Dietrich}\ \emph {et~al.}(2016)\citenamefont {Dietrich}, \citenamefont {Fiore}, \citenamefont {Thompson}, \citenamefont {Kamp},\ and\ \citenamefont {H{\"o}fling}}]{dietrich2016gaas}%
  \BibitemOpen
  \bibfield  {author} {\bibinfo {author} {\bibfnamefont {C.~P.}\ \bibnamefont {Dietrich}}, \bibinfo {author} {\bibfnamefont {A.}~\bibnamefont {Fiore}}, \bibinfo {author} {\bibfnamefont {M.~G.}\ \bibnamefont {Thompson}}, \bibinfo {author} {\bibfnamefont {M.}~\bibnamefont {Kamp}},\ and\ \bibinfo {author} {\bibfnamefont {S.}~\bibnamefont {H{\"o}fling}},\ }\bibfield  {title} {\bibinfo {title} {Gaas integrated quantum photonics: towards compact and multi-functional quantum photonic integrated circuits},\ }\href@noop {} {\bibfield  {journal} {\bibinfo  {journal} {Laser \& Photonics Reviews}\ }\textbf {\bibinfo {volume} {10}},\ \bibinfo {pages} {870} (\bibinfo {year} {2016})}\BibitemShut {NoStop}%
\bibitem [{\citenamefont {Uppu}\ \emph {et~al.}(2021)\citenamefont {Uppu}, \citenamefont {Midolo}, \citenamefont {Zhou}, \citenamefont {Carolan},\ and\ \citenamefont {Lodahl}}]{Uppu2021}%
  \BibitemOpen
  \bibfield  {author} {\bibinfo {author} {\bibfnamefont {R.}~\bibnamefont {Uppu}}, \bibinfo {author} {\bibfnamefont {L.}~\bibnamefont {Midolo}}, \bibinfo {author} {\bibfnamefont {X.}~\bibnamefont {Zhou}}, \bibinfo {author} {\bibfnamefont {J.}~\bibnamefont {Carolan}},\ and\ \bibinfo {author} {\bibfnamefont {P.}~\bibnamefont {Lodahl}},\ }\bibfield  {title} {\bibinfo {title} {Quantum-dot-based deterministic photon--emitter interfaces for scalable photonic quantum technology},\ }\href {https://doi.org/10.1038/s41565-021-00965-6} {\bibfield  {journal} {\bibinfo  {journal} {Nature Nanotechnology}\ }\textbf {\bibinfo {volume} {16}},\ \bibinfo {pages} {1308} (\bibinfo {year} {2021})}\BibitemShut {NoStop}%
\bibitem [{\citenamefont {Maring}\ \emph {et~al.}(2024)\citenamefont {Maring}, \citenamefont {Fyrillas}, \citenamefont {Pont}, \citenamefont {Ivanov}, \citenamefont {Stepanov}, \citenamefont {Margaria}, \citenamefont {Hease}, \citenamefont {Pishchagin}, \citenamefont {Lema{\^\i}tre}, \citenamefont {Sagnes} \emph {et~al.}}]{maring2024versatile}%
  \BibitemOpen
  \bibfield  {author} {\bibinfo {author} {\bibfnamefont {N.}~\bibnamefont {Maring}}, \bibinfo {author} {\bibfnamefont {A.}~\bibnamefont {Fyrillas}}, \bibinfo {author} {\bibfnamefont {M.}~\bibnamefont {Pont}}, \bibinfo {author} {\bibfnamefont {E.}~\bibnamefont {Ivanov}}, \bibinfo {author} {\bibfnamefont {P.}~\bibnamefont {Stepanov}}, \bibinfo {author} {\bibfnamefont {N.}~\bibnamefont {Margaria}}, \bibinfo {author} {\bibfnamefont {W.}~\bibnamefont {Hease}}, \bibinfo {author} {\bibfnamefont {A.}~\bibnamefont {Pishchagin}}, \bibinfo {author} {\bibfnamefont {A.}~\bibnamefont {Lema{\^\i}tre}}, \bibinfo {author} {\bibfnamefont {I.}~\bibnamefont {Sagnes}}, \emph {et~al.},\ }\bibfield  {title} {\bibinfo {title} {A versatile single-photon-based quantum computing platform},\ }\href@noop {} {\bibfield  {journal} {\bibinfo  {journal} {Nature Photonics}\ }\textbf {\bibinfo {volume} {18}},\ \bibinfo {pages} {603} (\bibinfo {year} {2024})}\BibitemShut {NoStop}%
\bibitem [{\citenamefont {Morrison}\ \emph {et~al.}(2023)\citenamefont {Morrison}, \citenamefont {Pousa}, \citenamefont {Graffitti}, \citenamefont {Koong}, \citenamefont {Barrow}, \citenamefont {Stoltz}, \citenamefont {Bouwmeester}, \citenamefont {Jeffers}, \citenamefont {Oi}, \citenamefont {Gerardot} \emph {et~al.}}]{morrison2023single}%
  \BibitemOpen
  \bibfield  {author} {\bibinfo {author} {\bibfnamefont {C.~L.}\ \bibnamefont {Morrison}}, \bibinfo {author} {\bibfnamefont {R.~G.}\ \bibnamefont {Pousa}}, \bibinfo {author} {\bibfnamefont {F.}~\bibnamefont {Graffitti}}, \bibinfo {author} {\bibfnamefont {Z.~X.}\ \bibnamefont {Koong}}, \bibinfo {author} {\bibfnamefont {P.}~\bibnamefont {Barrow}}, \bibinfo {author} {\bibfnamefont {N.~G.}\ \bibnamefont {Stoltz}}, \bibinfo {author} {\bibfnamefont {D.}~\bibnamefont {Bouwmeester}}, \bibinfo {author} {\bibfnamefont {J.}~\bibnamefont {Jeffers}}, \bibinfo {author} {\bibfnamefont {D.~K.}\ \bibnamefont {Oi}}, \bibinfo {author} {\bibfnamefont {B.~D.}\ \bibnamefont {Gerardot}}, \emph {et~al.},\ }\bibfield  {title} {\bibinfo {title} {Single-emitter quantum key distribution over 175 km of fibre with optimised finite key rates},\ }\href@noop {} {\bibfield  {journal} {\bibinfo  {journal} {Nature Communications}\ }\textbf {\bibinfo {volume} {14}},\ \bibinfo {pages} {3573} (\bibinfo {year} {2023})}\BibitemShut {NoStop}%
\bibitem [{\citenamefont {Zahidy}\ \emph {et~al.}(2024)\citenamefont {Zahidy}, \citenamefont {Mikkelsen}, \citenamefont {M{\"u}ller}, \citenamefont {Da~Lio}, \citenamefont {Krehbiel}, \citenamefont {Wang}, \citenamefont {Bart}, \citenamefont {Wieck}, \citenamefont {Ludwig}, \citenamefont {Galili} \emph {et~al.}}]{zahidy2024quantum}%
  \BibitemOpen
  \bibfield  {author} {\bibinfo {author} {\bibfnamefont {M.}~\bibnamefont {Zahidy}}, \bibinfo {author} {\bibfnamefont {M.~T.}\ \bibnamefont {Mikkelsen}}, \bibinfo {author} {\bibfnamefont {R.}~\bibnamefont {M{\"u}ller}}, \bibinfo {author} {\bibfnamefont {B.}~\bibnamefont {Da~Lio}}, \bibinfo {author} {\bibfnamefont {M.}~\bibnamefont {Krehbiel}}, \bibinfo {author} {\bibfnamefont {Y.}~\bibnamefont {Wang}}, \bibinfo {author} {\bibfnamefont {N.}~\bibnamefont {Bart}}, \bibinfo {author} {\bibfnamefont {A.~D.}\ \bibnamefont {Wieck}}, \bibinfo {author} {\bibfnamefont {A.}~\bibnamefont {Ludwig}}, \bibinfo {author} {\bibfnamefont {M.}~\bibnamefont {Galili}}, \emph {et~al.},\ }\bibfield  {title} {\bibinfo {title} {Quantum key distribution using deterministic single-photon sources over a field-installed fibre link},\ }\href@noop {} {\bibfield  {journal} {\bibinfo  {journal} {npj Quantum Information}\ }\textbf {\bibinfo {volume} {10}},\ \bibinfo {pages} {2} (\bibinfo {year} {2024})}\BibitemShut {NoStop}%
\bibitem [{\citenamefont {Wang}\ \emph {et~al.}(2023)\citenamefont {Wang}, \citenamefont {Faurby}, \citenamefont {Ruf}, \citenamefont {Sund}, \citenamefont {Nielsen}, \citenamefont {Volet}, \citenamefont {Heck}, \citenamefont {Bart}, \citenamefont {Wieck}, \citenamefont {Ludwig} \emph {et~al.}}]{wang2023deterministic}%
  \BibitemOpen
  \bibfield  {author} {\bibinfo {author} {\bibfnamefont {Y.}~\bibnamefont {Wang}}, \bibinfo {author} {\bibfnamefont {C.~F.}\ \bibnamefont {Faurby}}, \bibinfo {author} {\bibfnamefont {F.}~\bibnamefont {Ruf}}, \bibinfo {author} {\bibfnamefont {P.~I.}\ \bibnamefont {Sund}}, \bibinfo {author} {\bibfnamefont {K.}~\bibnamefont {Nielsen}}, \bibinfo {author} {\bibfnamefont {N.}~\bibnamefont {Volet}}, \bibinfo {author} {\bibfnamefont {M.~J.}\ \bibnamefont {Heck}}, \bibinfo {author} {\bibfnamefont {N.}~\bibnamefont {Bart}}, \bibinfo {author} {\bibfnamefont {A.~D.}\ \bibnamefont {Wieck}}, \bibinfo {author} {\bibfnamefont {A.}~\bibnamefont {Ludwig}}, \emph {et~al.},\ }\bibfield  {title} {\bibinfo {title} {Deterministic photon source interfaced with a programmable silicon-nitride integrated circuit},\ }\href@noop {} {\bibfield  {journal} {\bibinfo  {journal} {npj Quantum Information}\ }\textbf {\bibinfo {volume} {9}},\ \bibinfo {pages} {94} (\bibinfo {year} {2023})}\BibitemShut {NoStop}%
\bibitem [{\citenamefont {Uppu}\ \emph {et~al.}(2020)\citenamefont {Uppu}, \citenamefont {Pedersen}, \citenamefont {Wang}, \citenamefont {Olesen}, \citenamefont {Papon}, \citenamefont {Zhou}, \citenamefont {Midolo}, \citenamefont {Scholz}, \citenamefont {Wieck}, \citenamefont {Ludwig} \emph {et~al.}}]{uppu2020scalable}%
  \BibitemOpen
  \bibfield  {author} {\bibinfo {author} {\bibfnamefont {R.}~\bibnamefont {Uppu}}, \bibinfo {author} {\bibfnamefont {F.~T.}\ \bibnamefont {Pedersen}}, \bibinfo {author} {\bibfnamefont {Y.}~\bibnamefont {Wang}}, \bibinfo {author} {\bibfnamefont {C.~T.}\ \bibnamefont {Olesen}}, \bibinfo {author} {\bibfnamefont {C.}~\bibnamefont {Papon}}, \bibinfo {author} {\bibfnamefont {X.}~\bibnamefont {Zhou}}, \bibinfo {author} {\bibfnamefont {L.}~\bibnamefont {Midolo}}, \bibinfo {author} {\bibfnamefont {S.}~\bibnamefont {Scholz}}, \bibinfo {author} {\bibfnamefont {A.~D.}\ \bibnamefont {Wieck}}, \bibinfo {author} {\bibfnamefont {A.}~\bibnamefont {Ludwig}}, \emph {et~al.},\ }\bibfield  {title} {\bibinfo {title} {Scalable integrated single-photon source},\ }\href@noop {} {\bibfield  {journal} {\bibinfo  {journal} {Science advances}\ }\textbf {\bibinfo {volume} {6}},\ \bibinfo {pages} {eabc8268} (\bibinfo {year} {2020})}\BibitemShut {NoStop}%
\bibitem [{\citenamefont {Tomm}\ \emph {et~al.}(2021)\citenamefont {Tomm}, \citenamefont {Javadi}, \citenamefont {Antoniadis}, \citenamefont {Najer}, \citenamefont {L{\"o}bl}, \citenamefont {Korsch}, \citenamefont {Schott}, \citenamefont {Valentin}, \citenamefont {Wieck}, \citenamefont {Ludwig} \emph {et~al.}}]{tomm2021bright}%
  \BibitemOpen
  \bibfield  {author} {\bibinfo {author} {\bibfnamefont {N.}~\bibnamefont {Tomm}}, \bibinfo {author} {\bibfnamefont {A.}~\bibnamefont {Javadi}}, \bibinfo {author} {\bibfnamefont {N.~O.}\ \bibnamefont {Antoniadis}}, \bibinfo {author} {\bibfnamefont {D.}~\bibnamefont {Najer}}, \bibinfo {author} {\bibfnamefont {M.~C.}\ \bibnamefont {L{\"o}bl}}, \bibinfo {author} {\bibfnamefont {A.~R.}\ \bibnamefont {Korsch}}, \bibinfo {author} {\bibfnamefont {R.}~\bibnamefont {Schott}}, \bibinfo {author} {\bibfnamefont {S.~R.}\ \bibnamefont {Valentin}}, \bibinfo {author} {\bibfnamefont {A.~D.}\ \bibnamefont {Wieck}}, \bibinfo {author} {\bibfnamefont {A.}~\bibnamefont {Ludwig}}, \emph {et~al.},\ }\bibfield  {title} {\bibinfo {title} {A bright and fast source of coherent single photons},\ }\href@noop {} {\bibfield  {journal} {\bibinfo  {journal} {Nature Nanotechnology}\ }\textbf {\bibinfo {volume} {16}},\ \bibinfo {pages} {399} (\bibinfo {year} {2021})}\BibitemShut {NoStop}%
\bibitem [{\citenamefont {Papon}\ \emph {et~al.}(2023)\citenamefont {Papon}, \citenamefont {Wang}, \citenamefont {Uppu}, \citenamefont {Scholz}, \citenamefont {Wieck}, \citenamefont {Ludwig}, \citenamefont {Lodahl},\ and\ \citenamefont {Midolo}}]{papon2023independent}%
  \BibitemOpen
  \bibfield  {author} {\bibinfo {author} {\bibfnamefont {C.}~\bibnamefont {Papon}}, \bibinfo {author} {\bibfnamefont {Y.}~\bibnamefont {Wang}}, \bibinfo {author} {\bibfnamefont {R.}~\bibnamefont {Uppu}}, \bibinfo {author} {\bibfnamefont {S.}~\bibnamefont {Scholz}}, \bibinfo {author} {\bibfnamefont {A.~D.}\ \bibnamefont {Wieck}}, \bibinfo {author} {\bibfnamefont {A.}~\bibnamefont {Ludwig}}, \bibinfo {author} {\bibfnamefont {P.}~\bibnamefont {Lodahl}},\ and\ \bibinfo {author} {\bibfnamefont {L.}~\bibnamefont {Midolo}},\ }\bibfield  {title} {\bibinfo {title} {Independent operation of two waveguide-integrated quantum emitters},\ }\href@noop {} {\bibfield  {journal} {\bibinfo  {journal} {Physical Review Applied}\ }\textbf {\bibinfo {volume} {19}},\ \bibinfo {pages} {L061003} (\bibinfo {year} {2023})}\BibitemShut {NoStop}%
\bibitem [{\citenamefont {Sprengers}\ \emph {et~al.}(2011)\citenamefont {Sprengers}, \citenamefont {Gaggero}, \citenamefont {Sahin}, \citenamefont {Jahanmirinejad}, \citenamefont {Frucci}, \citenamefont {Mattioli}, \citenamefont {Leoni}, \citenamefont {Beetz}, \citenamefont {Lermer}, \citenamefont {Kamp} \emph {et~al.}}]{sprengers2011waveguide}%
  \BibitemOpen
  \bibfield  {author} {\bibinfo {author} {\bibfnamefont {J.}~\bibnamefont {Sprengers}}, \bibinfo {author} {\bibfnamefont {A.}~\bibnamefont {Gaggero}}, \bibinfo {author} {\bibfnamefont {D.}~\bibnamefont {Sahin}}, \bibinfo {author} {\bibfnamefont {S.}~\bibnamefont {Jahanmirinejad}}, \bibinfo {author} {\bibfnamefont {G.}~\bibnamefont {Frucci}}, \bibinfo {author} {\bibfnamefont {F.}~\bibnamefont {Mattioli}}, \bibinfo {author} {\bibfnamefont {R.}~\bibnamefont {Leoni}}, \bibinfo {author} {\bibfnamefont {J.}~\bibnamefont {Beetz}}, \bibinfo {author} {\bibfnamefont {M.}~\bibnamefont {Lermer}}, \bibinfo {author} {\bibfnamefont {M.}~\bibnamefont {Kamp}}, \emph {et~al.},\ }\bibfield  {title} {\bibinfo {title} {Waveguide superconducting single-photon detectors for integrated quantum photonic circuits},\ }\href@noop {} {\bibfield  {journal} {\bibinfo  {journal} {Applied Physics Letters}\ }\textbf {\bibinfo {volume} {99}} (\bibinfo {year} {2011})}\BibitemShut {NoStop}%
\bibitem [{\citenamefont {Reithmaier}\ \emph {et~al.}(2013)\citenamefont {Reithmaier}, \citenamefont {Senf}, \citenamefont {Lichtmannecker}, \citenamefont {Reichert}, \citenamefont {Flassig}, \citenamefont {Voss}, \citenamefont {Gross},\ and\ \citenamefont {Finley}}]{reithmaier2013optimisation}%
  \BibitemOpen
  \bibfield  {author} {\bibinfo {author} {\bibfnamefont {G.}~\bibnamefont {Reithmaier}}, \bibinfo {author} {\bibfnamefont {J.}~\bibnamefont {Senf}}, \bibinfo {author} {\bibfnamefont {S.}~\bibnamefont {Lichtmannecker}}, \bibinfo {author} {\bibfnamefont {T.}~\bibnamefont {Reichert}}, \bibinfo {author} {\bibfnamefont {F.}~\bibnamefont {Flassig}}, \bibinfo {author} {\bibfnamefont {A.}~\bibnamefont {Voss}}, \bibinfo {author} {\bibfnamefont {R.}~\bibnamefont {Gross}},\ and\ \bibinfo {author} {\bibfnamefont {J.}~\bibnamefont {Finley}},\ }\bibfield  {title} {\bibinfo {title} {Optimisation of nbn thin films on gaas substrates for in-situ single photon detection in structured photonic devices},\ }\href@noop {} {\bibfield  {journal} {\bibinfo  {journal} {Journal of Applied Physics}\ }\textbf {\bibinfo {volume} {113}} (\bibinfo {year} {2013})}\BibitemShut {NoStop}%
\bibitem [{\citenamefont {Bogaerts}\ \emph {et~al.}(2020)\citenamefont {Bogaerts}, \citenamefont {P{\'e}rez}, \citenamefont {Capmany}, \citenamefont {Miller}, \citenamefont {Poon}, \citenamefont {Englund}, \citenamefont {Morichetti},\ and\ \citenamefont {Melloni}}]{bogaerts2020programmable}%
  \BibitemOpen
  \bibfield  {author} {\bibinfo {author} {\bibfnamefont {W.}~\bibnamefont {Bogaerts}}, \bibinfo {author} {\bibfnamefont {D.}~\bibnamefont {P{\'e}rez}}, \bibinfo {author} {\bibfnamefont {J.}~\bibnamefont {Capmany}}, \bibinfo {author} {\bibfnamefont {D.~A.}\ \bibnamefont {Miller}}, \bibinfo {author} {\bibfnamefont {J.}~\bibnamefont {Poon}}, \bibinfo {author} {\bibfnamefont {D.}~\bibnamefont {Englund}}, \bibinfo {author} {\bibfnamefont {F.}~\bibnamefont {Morichetti}},\ and\ \bibinfo {author} {\bibfnamefont {A.}~\bibnamefont {Melloni}},\ }\bibfield  {title} {\bibinfo {title} {Programmable photonic circuits},\ }\href@noop {} {\bibfield  {journal} {\bibinfo  {journal} {Nature}\ }\textbf {\bibinfo {volume} {586}},\ \bibinfo {pages} {207} (\bibinfo {year} {2020})}\BibitemShut {NoStop}%
\bibitem [{\citenamefont {Harris}\ \emph {et~al.}(2014)\citenamefont {Harris}, \citenamefont {Ma}, \citenamefont {Mower}, \citenamefont {Baehr-Jones}, \citenamefont {Englund}, \citenamefont {Hochberg},\ and\ \citenamefont {Galland}}]{harris2014efficient}%
  \BibitemOpen
  \bibfield  {author} {\bibinfo {author} {\bibfnamefont {N.~C.}\ \bibnamefont {Harris}}, \bibinfo {author} {\bibfnamefont {Y.}~\bibnamefont {Ma}}, \bibinfo {author} {\bibfnamefont {J.}~\bibnamefont {Mower}}, \bibinfo {author} {\bibfnamefont {T.}~\bibnamefont {Baehr-Jones}}, \bibinfo {author} {\bibfnamefont {D.}~\bibnamefont {Englund}}, \bibinfo {author} {\bibfnamefont {M.}~\bibnamefont {Hochberg}},\ and\ \bibinfo {author} {\bibfnamefont {C.}~\bibnamefont {Galland}},\ }\bibfield  {title} {\bibinfo {title} {Efficient, compact and low loss thermo-optic phase shifter in silicon},\ }\href@noop {} {\bibfield  {journal} {\bibinfo  {journal} {Optics express}\ }\textbf {\bibinfo {volume} {22}},\ \bibinfo {pages} {10487} (\bibinfo {year} {2014})}\BibitemShut {NoStop}%
\bibitem [{\citenamefont {Nejadriahi}\ \emph {et~al.}(2021)\citenamefont {Nejadriahi}, \citenamefont {Pappert}, \citenamefont {Fainman},\ and\ \citenamefont {Yu}}]{nejadriahi2021efficient}%
  \BibitemOpen
  \bibfield  {author} {\bibinfo {author} {\bibfnamefont {H.}~\bibnamefont {Nejadriahi}}, \bibinfo {author} {\bibfnamefont {S.}~\bibnamefont {Pappert}}, \bibinfo {author} {\bibfnamefont {Y.}~\bibnamefont {Fainman}},\ and\ \bibinfo {author} {\bibfnamefont {P.}~\bibnamefont {Yu}},\ }\bibfield  {title} {\bibinfo {title} {Efficient and compact thermo-optic phase shifter in silicon-rich silicon nitride},\ }\href@noop {} {\bibfield  {journal} {\bibinfo  {journal} {Optics letters}\ }\textbf {\bibinfo {volume} {46}},\ \bibinfo {pages} {4646} (\bibinfo {year} {2021})}\BibitemShut {NoStop}%
\bibitem [{\citenamefont {Wood}\ \emph {et~al.}(1984)\citenamefont {Wood}, \citenamefont {Burrus}, \citenamefont {Miller}, \citenamefont {Chemla}, \citenamefont {Damen}, \citenamefont {Gossard},\ and\ \citenamefont {Wiegmann}}]{wood1984high}%
  \BibitemOpen
  \bibfield  {author} {\bibinfo {author} {\bibfnamefont {T.}~\bibnamefont {Wood}}, \bibinfo {author} {\bibfnamefont {C.}~\bibnamefont {Burrus}}, \bibinfo {author} {\bibfnamefont {D.}~\bibnamefont {Miller}}, \bibinfo {author} {\bibfnamefont {D.}~\bibnamefont {Chemla}}, \bibinfo {author} {\bibfnamefont {T.}~\bibnamefont {Damen}}, \bibinfo {author} {\bibfnamefont {A.}~\bibnamefont {Gossard}},\ and\ \bibinfo {author} {\bibfnamefont {W.}~\bibnamefont {Wiegmann}},\ }\bibfield  {title} {\bibinfo {title} {High-speed optical modulation with gaas/gaalas quantum wells in ap-i-n diode structure},\ }\href@noop {} {\bibfield  {journal} {\bibinfo  {journal} {Applied Physics Letters}\ }\textbf {\bibinfo {volume} {44}},\ \bibinfo {pages} {16} (\bibinfo {year} {1984})}\BibitemShut {NoStop}%
\bibitem [{\citenamefont {Thomaschewski}\ and\ \citenamefont {Bozhevolnyi}(2022)}]{thomaschewski2022pockels}%
  \BibitemOpen
  \bibfield  {author} {\bibinfo {author} {\bibfnamefont {M.}~\bibnamefont {Thomaschewski}}\ and\ \bibinfo {author} {\bibfnamefont {S.}~\bibnamefont {Bozhevolnyi}},\ }\bibfield  {title} {\bibinfo {title} {Pockels modulation in integrated nanophotonics},\ }\href@noop {} {\bibfield  {journal} {\bibinfo  {journal} {Applied Physics Reviews}\ }\textbf {\bibinfo {volume} {9}} (\bibinfo {year} {2022})}\BibitemShut {NoStop}%
\bibitem [{\citenamefont {Midolo}\ \emph {et~al.}(2017)\citenamefont {Midolo}, \citenamefont {Hansen}, \citenamefont {Zhang}, \citenamefont {Papon}, \citenamefont {Schott}, \citenamefont {Ludwig}, \citenamefont {Wieck}, \citenamefont {Lodahl},\ and\ \citenamefont {Stobbe}}]{midolo2017electro}%
  \BibitemOpen
  \bibfield  {author} {\bibinfo {author} {\bibfnamefont {L.}~\bibnamefont {Midolo}}, \bibinfo {author} {\bibfnamefont {S.~L.}\ \bibnamefont {Hansen}}, \bibinfo {author} {\bibfnamefont {W.}~\bibnamefont {Zhang}}, \bibinfo {author} {\bibfnamefont {C.}~\bibnamefont {Papon}}, \bibinfo {author} {\bibfnamefont {R.}~\bibnamefont {Schott}}, \bibinfo {author} {\bibfnamefont {A.}~\bibnamefont {Ludwig}}, \bibinfo {author} {\bibfnamefont {A.~D.}\ \bibnamefont {Wieck}}, \bibinfo {author} {\bibfnamefont {P.}~\bibnamefont {Lodahl}},\ and\ \bibinfo {author} {\bibfnamefont {S.}~\bibnamefont {Stobbe}},\ }\bibfield  {title} {\bibinfo {title} {Electro-optic routing of photons from a single quantum dot in photonic integrated circuits},\ }\href@noop {} {\bibfield  {journal} {\bibinfo  {journal} {Optics Express}\ }\textbf {\bibinfo {volume} {25}},\ \bibinfo {pages} {33514} (\bibinfo {year} {2017})}\BibitemShut {NoStop}%
\bibitem [{\citenamefont {Wang}\ \emph {et~al.}(2021)\citenamefont {Wang}, \citenamefont {Uppu}, \citenamefont {Zhou}, \citenamefont {Papon}, \citenamefont {Scholz}, \citenamefont {Wieck}, \citenamefont {Ludwig}, \citenamefont {Lodahl},\ and\ \citenamefont {Midolo}}]{wang2021electroabsorption}%
  \BibitemOpen
  \bibfield  {author} {\bibinfo {author} {\bibfnamefont {Y.}~\bibnamefont {Wang}}, \bibinfo {author} {\bibfnamefont {R.}~\bibnamefont {Uppu}}, \bibinfo {author} {\bibfnamefont {X.}~\bibnamefont {Zhou}}, \bibinfo {author} {\bibfnamefont {C.}~\bibnamefont {Papon}}, \bibinfo {author} {\bibfnamefont {S.}~\bibnamefont {Scholz}}, \bibinfo {author} {\bibfnamefont {A.~D.}\ \bibnamefont {Wieck}}, \bibinfo {author} {\bibfnamefont {A.}~\bibnamefont {Ludwig}}, \bibinfo {author} {\bibfnamefont {P.}~\bibnamefont {Lodahl}},\ and\ \bibinfo {author} {\bibfnamefont {L.}~\bibnamefont {Midolo}},\ }\bibfield  {title} {\bibinfo {title} {Electroabsorption in gated gaas nanophotonic waveguides},\ }\href@noop {} {\bibfield  {journal} {\bibinfo  {journal} {Applied Physics Letters}\ }\textbf {\bibinfo {volume} {118}} (\bibinfo {year} {2021})}\BibitemShut {NoStop}%
\bibitem [{\citenamefont {Midolo}\ \emph {et~al.}(2018)\citenamefont {Midolo}, \citenamefont {Schliesser},\ and\ \citenamefont {Fiore}}]{midolo2018nano}%
  \BibitemOpen
  \bibfield  {author} {\bibinfo {author} {\bibfnamefont {L.}~\bibnamefont {Midolo}}, \bibinfo {author} {\bibfnamefont {A.}~\bibnamefont {Schliesser}},\ and\ \bibinfo {author} {\bibfnamefont {A.}~\bibnamefont {Fiore}},\ }\bibfield  {title} {\bibinfo {title} {Nano-opto-electro-mechanical systems},\ }\href@noop {} {\bibfield  {journal} {\bibinfo  {journal} {Nature nanotechnology}\ }\textbf {\bibinfo {volume} {13}},\ \bibinfo {pages} {11} (\bibinfo {year} {2018})}\BibitemShut {NoStop}%
\bibitem [{\citenamefont {Papon}\ \emph {et~al.}(2019)\citenamefont {Papon}, \citenamefont {Zhou}, \citenamefont {Thyrrestrup}, \citenamefont {Liu}, \citenamefont {Stobbe}, \citenamefont {Schott}, \citenamefont {Wieck}, \citenamefont {Ludwig}, \citenamefont {Lodahl},\ and\ \citenamefont {Midolo}}]{papon2019nanomechanical}%
  \BibitemOpen
  \bibfield  {author} {\bibinfo {author} {\bibfnamefont {C.}~\bibnamefont {Papon}}, \bibinfo {author} {\bibfnamefont {X.}~\bibnamefont {Zhou}}, \bibinfo {author} {\bibfnamefont {H.}~\bibnamefont {Thyrrestrup}}, \bibinfo {author} {\bibfnamefont {Z.}~\bibnamefont {Liu}}, \bibinfo {author} {\bibfnamefont {S.}~\bibnamefont {Stobbe}}, \bibinfo {author} {\bibfnamefont {R.}~\bibnamefont {Schott}}, \bibinfo {author} {\bibfnamefont {A.~D.}\ \bibnamefont {Wieck}}, \bibinfo {author} {\bibfnamefont {A.}~\bibnamefont {Ludwig}}, \bibinfo {author} {\bibfnamefont {P.}~\bibnamefont {Lodahl}},\ and\ \bibinfo {author} {\bibfnamefont {L.}~\bibnamefont {Midolo}},\ }\bibfield  {title} {\bibinfo {title} {Nanomechanical single-photon routing},\ }\href@noop {} {\bibfield  {journal} {\bibinfo  {journal} {Optica}\ }\textbf {\bibinfo {volume} {6}},\ \bibinfo {pages} {524} (\bibinfo {year} {2019})}\BibitemShut {NoStop}%
\bibitem [{\citenamefont {Van~Acoleyen}\ \emph {et~al.}(2012)\citenamefont {Van~Acoleyen}, \citenamefont {Roels}, \citenamefont {Mechet}, \citenamefont {Claes}, \citenamefont {Van~Thourhout},\ and\ \citenamefont {Baets}}]{van2012ultracompact}%
  \BibitemOpen
  \bibfield  {author} {\bibinfo {author} {\bibfnamefont {K.}~\bibnamefont {Van~Acoleyen}}, \bibinfo {author} {\bibfnamefont {J.}~\bibnamefont {Roels}}, \bibinfo {author} {\bibfnamefont {P.}~\bibnamefont {Mechet}}, \bibinfo {author} {\bibfnamefont {T.}~\bibnamefont {Claes}}, \bibinfo {author} {\bibfnamefont {D.}~\bibnamefont {Van~Thourhout}},\ and\ \bibinfo {author} {\bibfnamefont {R.}~\bibnamefont {Baets}},\ }\bibfield  {title} {\bibinfo {title} {Ultracompact phase modulator based on a cascade of nems-operated slot waveguides fabricated in silicon-on-insulator},\ }\href@noop {} {\bibfield  {journal} {\bibinfo  {journal} {IEEE Photonics Journal}\ }\textbf {\bibinfo {volume} {4}},\ \bibinfo {pages} {779} (\bibinfo {year} {2012})}\BibitemShut {NoStop}%
\bibitem [{\citenamefont {Haffner}\ \emph {et~al.}(2019)\citenamefont {Haffner}, \citenamefont {Joerg}, \citenamefont {Doderer}, \citenamefont {Mayor}, \citenamefont {Chelladurai}, \citenamefont {Fedoryshyn}, \citenamefont {Roman}, \citenamefont {Mazur}, \citenamefont {Burla}, \citenamefont {Lezec} \emph {et~al.}}]{haffner2019nano}%
  \BibitemOpen
  \bibfield  {author} {\bibinfo {author} {\bibfnamefont {C.}~\bibnamefont {Haffner}}, \bibinfo {author} {\bibfnamefont {A.}~\bibnamefont {Joerg}}, \bibinfo {author} {\bibfnamefont {M.}~\bibnamefont {Doderer}}, \bibinfo {author} {\bibfnamefont {F.}~\bibnamefont {Mayor}}, \bibinfo {author} {\bibfnamefont {D.}~\bibnamefont {Chelladurai}}, \bibinfo {author} {\bibfnamefont {Y.}~\bibnamefont {Fedoryshyn}}, \bibinfo {author} {\bibfnamefont {C.~I.}\ \bibnamefont {Roman}}, \bibinfo {author} {\bibfnamefont {M.}~\bibnamefont {Mazur}}, \bibinfo {author} {\bibfnamefont {M.}~\bibnamefont {Burla}}, \bibinfo {author} {\bibfnamefont {H.~J.}\ \bibnamefont {Lezec}}, \emph {et~al.},\ }\bibfield  {title} {\bibinfo {title} {Nano--opto-electro-mechanical switches operated at cmos-level voltages},\ }\href@noop {} {\bibfield  {journal} {\bibinfo  {journal} {Science}\ }\textbf {\bibinfo {volume} {366}},\ \bibinfo {pages} {860} (\bibinfo {year} {2019})}\BibitemShut {NoStop}%
\bibitem [{\citenamefont {Edinger}\ \emph {et~al.}(2021)\citenamefont {Edinger}, \citenamefont {Takabayashi}, \citenamefont {Errando-Herranz}, \citenamefont {Khan}, \citenamefont {Sattari}, \citenamefont {Verheyen}, \citenamefont {Bogaerts}, \citenamefont {Quack},\ and\ \citenamefont {Gylfason}}]{edinger2021silicon}%
  \BibitemOpen
  \bibfield  {author} {\bibinfo {author} {\bibfnamefont {P.}~\bibnamefont {Edinger}}, \bibinfo {author} {\bibfnamefont {A.~Y.}\ \bibnamefont {Takabayashi}}, \bibinfo {author} {\bibfnamefont {C.}~\bibnamefont {Errando-Herranz}}, \bibinfo {author} {\bibfnamefont {U.}~\bibnamefont {Khan}}, \bibinfo {author} {\bibfnamefont {H.}~\bibnamefont {Sattari}}, \bibinfo {author} {\bibfnamefont {P.}~\bibnamefont {Verheyen}}, \bibinfo {author} {\bibfnamefont {W.}~\bibnamefont {Bogaerts}}, \bibinfo {author} {\bibfnamefont {N.}~\bibnamefont {Quack}},\ and\ \bibinfo {author} {\bibfnamefont {K.~B.}\ \bibnamefont {Gylfason}},\ }\bibfield  {title} {\bibinfo {title} {Silicon photonic microelectromechanical phase shifters for scalable programmable photonics},\ }\href@noop {} {\bibfield  {journal} {\bibinfo  {journal} {Optics Letters}\ }\textbf {\bibinfo {volume} {46}},\ \bibinfo {pages} {5671} (\bibinfo {year} {2021})}\BibitemShut {NoStop}%
\bibitem [{\citenamefont {Baghdadi}\ \emph {et~al.}(2021)\citenamefont {Baghdadi}, \citenamefont {Gould}, \citenamefont {Gupta}, \citenamefont {Tymchenko}, \citenamefont {Bunandar}, \citenamefont {Ramey},\ and\ \citenamefont {Harris}}]{baghdadi2021dual}%
  \BibitemOpen
  \bibfield  {author} {\bibinfo {author} {\bibfnamefont {R.}~\bibnamefont {Baghdadi}}, \bibinfo {author} {\bibfnamefont {M.}~\bibnamefont {Gould}}, \bibinfo {author} {\bibfnamefont {S.}~\bibnamefont {Gupta}}, \bibinfo {author} {\bibfnamefont {M.}~\bibnamefont {Tymchenko}}, \bibinfo {author} {\bibfnamefont {D.}~\bibnamefont {Bunandar}}, \bibinfo {author} {\bibfnamefont {C.}~\bibnamefont {Ramey}},\ and\ \bibinfo {author} {\bibfnamefont {N.~C.}\ \bibnamefont {Harris}},\ }\bibfield  {title} {\bibinfo {title} {Dual slot-mode noem phase shifter},\ }\href@noop {} {\bibfield  {journal} {\bibinfo  {journal} {Optics Express}\ }\textbf {\bibinfo {volume} {29}},\ \bibinfo {pages} {19113} (\bibinfo {year} {2021})}\BibitemShut {NoStop}%
\bibitem [{\citenamefont {Almeida}\ \emph {et~al.}(2004)\citenamefont {Almeida}, \citenamefont {Xu}, \citenamefont {Barrios},\ and\ \citenamefont {Lipson}}]{almeida2004guiding}%
  \BibitemOpen
  \bibfield  {author} {\bibinfo {author} {\bibfnamefont {V.~R.}\ \bibnamefont {Almeida}}, \bibinfo {author} {\bibfnamefont {Q.}~\bibnamefont {Xu}}, \bibinfo {author} {\bibfnamefont {C.~A.}\ \bibnamefont {Barrios}},\ and\ \bibinfo {author} {\bibfnamefont {M.}~\bibnamefont {Lipson}},\ }\bibfield  {title} {\bibinfo {title} {Guiding and confining light in void nanostructure},\ }\href@noop {} {\bibfield  {journal} {\bibinfo  {journal} {Optics letters}\ }\textbf {\bibinfo {volume} {29}},\ \bibinfo {pages} {1209} (\bibinfo {year} {2004})}\BibitemShut {NoStop}%
\bibitem [{\citenamefont {Grottke}\ \emph {et~al.}(2021)\citenamefont {Grottke}, \citenamefont {Hartmann}, \citenamefont {Schuck},\ and\ \citenamefont {Pernice}}]{grottke2021optoelectromechanical}%
  \BibitemOpen
  \bibfield  {author} {\bibinfo {author} {\bibfnamefont {T.}~\bibnamefont {Grottke}}, \bibinfo {author} {\bibfnamefont {W.}~\bibnamefont {Hartmann}}, \bibinfo {author} {\bibfnamefont {C.}~\bibnamefont {Schuck}},\ and\ \bibinfo {author} {\bibfnamefont {W.~H.}\ \bibnamefont {Pernice}},\ }\bibfield  {title} {\bibinfo {title} {Optoelectromechanical phase shifter with low insertion loss and a 13$\pi$ tuning range},\ }\href@noop {} {\bibfield  {journal} {\bibinfo  {journal} {Optics Express}\ }\textbf {\bibinfo {volume} {29}},\ \bibinfo {pages} {5525} (\bibinfo {year} {2021})}\BibitemShut {NoStop}%
\bibitem [{\citenamefont {Beutel}\ \emph {et~al.}(2022)\citenamefont {Beutel}, \citenamefont {Grottke}, \citenamefont {Wolff}, \citenamefont {Schuck},\ and\ \citenamefont {Pernice}}]{beutel2022cryo}%
  \BibitemOpen
  \bibfield  {author} {\bibinfo {author} {\bibfnamefont {F.}~\bibnamefont {Beutel}}, \bibinfo {author} {\bibfnamefont {T.}~\bibnamefont {Grottke}}, \bibinfo {author} {\bibfnamefont {M.~A.}\ \bibnamefont {Wolff}}, \bibinfo {author} {\bibfnamefont {C.}~\bibnamefont {Schuck}},\ and\ \bibinfo {author} {\bibfnamefont {W.~H.}\ \bibnamefont {Pernice}},\ }\bibfield  {title} {\bibinfo {title} {Cryo-compatible opto-mechanical low-voltage phase-modulator integrated with superconducting single-photon detectors},\ }\href@noop {} {\bibfield  {journal} {\bibinfo  {journal} {Optics Express}\ }\textbf {\bibinfo {volume} {30}},\ \bibinfo {pages} {30066} (\bibinfo {year} {2022})}\BibitemShut {NoStop}%
\bibitem [{\citenamefont {Legtenberg}\ \emph {et~al.}(1996)\citenamefont {Legtenberg}, \citenamefont {Groeneveld},\ and\ \citenamefont {Elwenspoek}}]{legtenberg1996comb}%
  \BibitemOpen
  \bibfield  {author} {\bibinfo {author} {\bibfnamefont {R.}~\bibnamefont {Legtenberg}}, \bibinfo {author} {\bibfnamefont {A.}~\bibnamefont {Groeneveld}},\ and\ \bibinfo {author} {\bibfnamefont {M.}~\bibnamefont {Elwenspoek}},\ }\bibfield  {title} {\bibinfo {title} {Comb-drive actuators for large displacements},\ }\href@noop {} {\bibfield  {journal} {\bibinfo  {journal} {Journal of Micromechanics and microengineering}\ }\textbf {\bibinfo {volume} {6}},\ \bibinfo {pages} {320} (\bibinfo {year} {1996})}\BibitemShut {NoStop}%
\bibitem [{\citenamefont {Zhou}\ \emph {et~al.}(2018)\citenamefont {Zhou}, \citenamefont {Kulkova}, \citenamefont {Lund-Hansen}, \citenamefont {Hansen}, \citenamefont {Lodahl},\ and\ \citenamefont {Midolo}}]{zhou2018high}%
  \BibitemOpen
  \bibfield  {author} {\bibinfo {author} {\bibfnamefont {X.}~\bibnamefont {Zhou}}, \bibinfo {author} {\bibfnamefont {I.}~\bibnamefont {Kulkova}}, \bibinfo {author} {\bibfnamefont {T.}~\bibnamefont {Lund-Hansen}}, \bibinfo {author} {\bibfnamefont {S.~L.}\ \bibnamefont {Hansen}}, \bibinfo {author} {\bibfnamefont {P.}~\bibnamefont {Lodahl}},\ and\ \bibinfo {author} {\bibfnamefont {L.}~\bibnamefont {Midolo}},\ }\bibfield  {title} {\bibinfo {title} {High-efficiency shallow-etched grating on gaas membranes for quantum photonic applications},\ }\href@noop {} {\bibfield  {journal} {\bibinfo  {journal} {Applied Physics Letters}\ }\textbf {\bibinfo {volume} {113}} (\bibinfo {year} {2018})}\BibitemShut {NoStop}%
\bibitem [{\citenamefont {Noel}(2016)}]{Noel2016GoldMembranes}%
  \BibitemOpen
  \bibfield  {author} {\bibinfo {author} {\bibfnamefont {J.~G.}\ \bibnamefont {Noel}},\ }\bibfield  {title} {\bibinfo {title} {Review of the properties of gold material for mems membrane applications},\ }\href {https://doi.org/https://doi.org/10.1049/iet-cds.2015.0094} {\bibfield  {journal} {\bibinfo  {journal} {IET Circuits, Devices and Systems}\ }\textbf {\bibinfo {volume} {10}},\ \bibinfo {pages} {156} (\bibinfo {year} {2016})}\BibitemShut {NoStop}%
\bibitem [{\citenamefont {Guha}\ \emph {et~al.}(2017)\citenamefont {Guha}, \citenamefont {Marsault}, \citenamefont {Cadiz}, \citenamefont {Morgenroth}, \citenamefont {Ulin}, \citenamefont {Berkovitz}, \citenamefont {Lema\^{i}tre}, \citenamefont {Gomez}, \citenamefont {Amo}, \citenamefont {Combri\'{e}}, \citenamefont {G\'{e}rard}, \citenamefont {Leo},\ and\ \citenamefont {Favero}}]{Guha:17}%
  \BibitemOpen
  \bibfield  {author} {\bibinfo {author} {\bibfnamefont {B.}~\bibnamefont {Guha}}, \bibinfo {author} {\bibfnamefont {F.}~\bibnamefont {Marsault}}, \bibinfo {author} {\bibfnamefont {F.}~\bibnamefont {Cadiz}}, \bibinfo {author} {\bibfnamefont {L.}~\bibnamefont {Morgenroth}}, \bibinfo {author} {\bibfnamefont {V.}~\bibnamefont {Ulin}}, \bibinfo {author} {\bibfnamefont {V.}~\bibnamefont {Berkovitz}}, \bibinfo {author} {\bibfnamefont {A.}~\bibnamefont {Lema\^{i}tre}}, \bibinfo {author} {\bibfnamefont {C.}~\bibnamefont {Gomez}}, \bibinfo {author} {\bibfnamefont {A.}~\bibnamefont {Amo}}, \bibinfo {author} {\bibfnamefont {S.}~\bibnamefont {Combri\'{e}}}, \bibinfo {author} {\bibfnamefont {B.}~\bibnamefont {G\'{e}rard}}, \bibinfo {author} {\bibfnamefont {G.}~\bibnamefont {Leo}},\ and\ \bibinfo {author} {\bibfnamefont {I.}~\bibnamefont {Favero}},\ }\bibfield  {title} {\bibinfo {title} {Surface-enhanced gallium arsenide photonic resonator with quality factor of 6 x $10^6$},\ }\href
  {https://doi.org/10.1364/OPTICA.4.000218} {\bibfield  {journal} {\bibinfo  {journal} {Optica}\ }\textbf {\bibinfo {volume} {4}},\ \bibinfo {pages} {218} (\bibinfo {year} {2017})}\BibitemShut {NoStop}%
\bibitem [{\citenamefont {Reck}\ \emph {et~al.}(1994)\citenamefont {Reck}, \citenamefont {Zeilinger}, \citenamefont {Bernstein},\ and\ \citenamefont {Bertani}}]{reck1994experimental}%
  \BibitemOpen
  \bibfield  {author} {\bibinfo {author} {\bibfnamefont {M.}~\bibnamefont {Reck}}, \bibinfo {author} {\bibfnamefont {A.}~\bibnamefont {Zeilinger}}, \bibinfo {author} {\bibfnamefont {H.~J.}\ \bibnamefont {Bernstein}},\ and\ \bibinfo {author} {\bibfnamefont {P.}~\bibnamefont {Bertani}},\ }\bibfield  {title} {\bibinfo {title} {Experimental realization of any discrete unitary operator},\ }\href@noop {} {\bibfield  {journal} {\bibinfo  {journal} {Physical review letters}\ }\textbf {\bibinfo {volume} {73}},\ \bibinfo {pages} {58} (\bibinfo {year} {1994})}\BibitemShut {NoStop}%
\bibitem [{\citenamefont {Taballione}\ \emph {et~al.}(2023)\citenamefont {Taballione}, \citenamefont {Anguita}, \citenamefont {de~Goede}, \citenamefont {Venderbosch}, \citenamefont {Kassenberg}, \citenamefont {Snijders}, \citenamefont {Kannan}, \citenamefont {Vleeshouwers}, \citenamefont {Smith}, \citenamefont {Epping} \emph {et~al.}}]{taballione202320}%
  \BibitemOpen
  \bibfield  {author} {\bibinfo {author} {\bibfnamefont {C.}~\bibnamefont {Taballione}}, \bibinfo {author} {\bibfnamefont {M.~C.}\ \bibnamefont {Anguita}}, \bibinfo {author} {\bibfnamefont {M.}~\bibnamefont {de~Goede}}, \bibinfo {author} {\bibfnamefont {P.}~\bibnamefont {Venderbosch}}, \bibinfo {author} {\bibfnamefont {B.}~\bibnamefont {Kassenberg}}, \bibinfo {author} {\bibfnamefont {H.}~\bibnamefont {Snijders}}, \bibinfo {author} {\bibfnamefont {N.}~\bibnamefont {Kannan}}, \bibinfo {author} {\bibfnamefont {W.~L.}\ \bibnamefont {Vleeshouwers}}, \bibinfo {author} {\bibfnamefont {D.}~\bibnamefont {Smith}}, \bibinfo {author} {\bibfnamefont {J.~P.}\ \bibnamefont {Epping}}, \emph {et~al.},\ }\bibfield  {title} {\bibinfo {title} {20-mode universal quantum photonic processor},\ }\href@noop {} {\bibfield  {journal} {\bibinfo  {journal} {Quantum}\ }\textbf {\bibinfo {volume} {7}},\ \bibinfo {pages} {1071} (\bibinfo {year} {2023})}\BibitemShut {NoStop}%
\bibitem [{\citenamefont {Sparrow}\ \emph {et~al.}(2018)\citenamefont {Sparrow}, \citenamefont {Mart{\'\i}n-L{\'o}pez}, \citenamefont {Maraviglia}, \citenamefont {Neville}, \citenamefont {Harrold}, \citenamefont {Carolan}, \citenamefont {Joglekar}, \citenamefont {Hashimoto}, \citenamefont {Matsuda}, \citenamefont {O’Brien} \emph {et~al.}}]{sparrow2018simulating}%
  \BibitemOpen
  \bibfield  {author} {\bibinfo {author} {\bibfnamefont {C.}~\bibnamefont {Sparrow}}, \bibinfo {author} {\bibfnamefont {E.}~\bibnamefont {Mart{\'\i}n-L{\'o}pez}}, \bibinfo {author} {\bibfnamefont {N.}~\bibnamefont {Maraviglia}}, \bibinfo {author} {\bibfnamefont {A.}~\bibnamefont {Neville}}, \bibinfo {author} {\bibfnamefont {C.}~\bibnamefont {Harrold}}, \bibinfo {author} {\bibfnamefont {J.}~\bibnamefont {Carolan}}, \bibinfo {author} {\bibfnamefont {Y.~N.}\ \bibnamefont {Joglekar}}, \bibinfo {author} {\bibfnamefont {T.}~\bibnamefont {Hashimoto}}, \bibinfo {author} {\bibfnamefont {N.}~\bibnamefont {Matsuda}}, \bibinfo {author} {\bibfnamefont {J.~L.}\ \bibnamefont {O’Brien}}, \emph {et~al.},\ }\bibfield  {title} {\bibinfo {title} {Simulating the vibrational quantum dynamics of molecules using photonics},\ }\href@noop {} {\bibfield  {journal} {\bibinfo  {journal} {Nature}\ }\textbf {\bibinfo {volume} {557}},\ \bibinfo {pages} {660} (\bibinfo {year} {2018})}\BibitemShut {NoStop}%
\bibitem [{\citenamefont {Bandyopadhyay}\ \emph {et~al.}(2024)\citenamefont {Bandyopadhyay}, \citenamefont {Sludds}, \citenamefont {Krastanov}, \citenamefont {Hamerly}, \citenamefont {Harris}, \citenamefont {Bunandar}, \citenamefont {Streshinsky}, \citenamefont {Hochberg},\ and\ \citenamefont {Englund}}]{bandyopadhyay2024single}%
  \BibitemOpen
  \bibfield  {author} {\bibinfo {author} {\bibfnamefont {S.}~\bibnamefont {Bandyopadhyay}}, \bibinfo {author} {\bibfnamefont {A.}~\bibnamefont {Sludds}}, \bibinfo {author} {\bibfnamefont {S.}~\bibnamefont {Krastanov}}, \bibinfo {author} {\bibfnamefont {R.}~\bibnamefont {Hamerly}}, \bibinfo {author} {\bibfnamefont {N.}~\bibnamefont {Harris}}, \bibinfo {author} {\bibfnamefont {D.}~\bibnamefont {Bunandar}}, \bibinfo {author} {\bibfnamefont {M.}~\bibnamefont {Streshinsky}}, \bibinfo {author} {\bibfnamefont {M.}~\bibnamefont {Hochberg}},\ and\ \bibinfo {author} {\bibfnamefont {D.}~\bibnamefont {Englund}},\ }\bibfield  {title} {\bibinfo {title} {Single-chip photonic deep neural network with forward-only training},\ }\href@noop {} {\bibfield  {journal} {\bibinfo  {journal} {Nature Photonics}\ }\textbf {\bibinfo {volume} {18}},\ \bibinfo {pages} {1335} (\bibinfo {year} {2024})}\BibitemShut {NoStop}%
\bibitem [{\citenamefont {Qvotrup}\ \emph {et~al.}(2025)\citenamefont {Qvotrup}, \citenamefont {Wang}, \citenamefont {Albrechtsen}, \citenamefont {Thomas}, \citenamefont {Liu}, \citenamefont {Scholz}, \citenamefont {Ludwig},\ and\ \citenamefont {Midolo}}]{companion}%
  \BibitemOpen
  \bibfield  {author} {\bibinfo {author} {\bibfnamefont {C.}~\bibnamefont {Qvotrup}}, \bibinfo {author} {\bibfnamefont {Y.}~\bibnamefont {Wang}}, \bibinfo {author} {\bibfnamefont {M.}~\bibnamefont {Albrechtsen}}, \bibinfo {author} {\bibfnamefont {R.}~\bibnamefont {Thomas}}, \bibinfo {author} {\bibfnamefont {Z.}~\bibnamefont {Liu}}, \bibinfo {author} {\bibfnamefont {S.}~\bibnamefont {Scholz}}, \bibinfo {author} {\bibfnamefont {A.}~\bibnamefont {Ludwig}},\ and\ \bibinfo {author} {\bibfnamefont {L.}~\bibnamefont {Midolo}},\ }\bibfield  {title} {\bibinfo {title} {Controlling emitter-field coupling in waveguides with nanomechanical phase shifters},\ }\href@noop {} {\bibfield  {journal} {\bibinfo  {journal} {Companion paper}\ } (\bibinfo {year} {2025})}\BibitemShut {NoStop}%
\bibitem [{\citenamefont {Tiranov}\ \emph {et~al.}(2023)\citenamefont {Tiranov}, \citenamefont {Angelopoulou}, \citenamefont {van Diepen}, \citenamefont {Schrinski}, \citenamefont {Sandberg}, \citenamefont {Wang}, \citenamefont {Midolo}, \citenamefont {Scholz}, \citenamefont {Wieck}, \citenamefont {Ludwig} \emph {et~al.}}]{tiranov2023collective}%
  \BibitemOpen
  \bibfield  {author} {\bibinfo {author} {\bibfnamefont {A.}~\bibnamefont {Tiranov}}, \bibinfo {author} {\bibfnamefont {V.}~\bibnamefont {Angelopoulou}}, \bibinfo {author} {\bibfnamefont {C.~J.}\ \bibnamefont {van Diepen}}, \bibinfo {author} {\bibfnamefont {B.}~\bibnamefont {Schrinski}}, \bibinfo {author} {\bibfnamefont {O.~A.~D.}\ \bibnamefont {Sandberg}}, \bibinfo {author} {\bibfnamefont {Y.}~\bibnamefont {Wang}}, \bibinfo {author} {\bibfnamefont {L.}~\bibnamefont {Midolo}}, \bibinfo {author} {\bibfnamefont {S.}~\bibnamefont {Scholz}}, \bibinfo {author} {\bibfnamefont {A.~D.}\ \bibnamefont {Wieck}}, \bibinfo {author} {\bibfnamefont {A.}~\bibnamefont {Ludwig}}, \emph {et~al.},\ }\bibfield  {title} {\bibinfo {title} {Collective super-and subradiant dynamics between distant optical quantum emitters},\ }\href@noop {} {\bibfield  {journal} {\bibinfo  {journal} {Science}\ }\textbf {\bibinfo {volume} {379}},\ \bibinfo {pages} {389} (\bibinfo {year} {2023})}\BibitemShut {NoStop}%
\end{thebibliography}%

\end{document}